\documentclass[10pt,conference]{IEEEtran}

\makeatletter
\def\ps@headings{%
\def\@oddhead{\mbox{}\scriptsize\rightmark \hfil \thepage}%
\def\@evenhead{\scriptsize\thepage \hfil \leftmark\mbox{}}%
\def\@oddfoot{}%
\def\@evenfoot{}}
\makeatother
\usepackage{amsfonts}
\usepackage{mathrsfs}
\usepackage{amsfonts}
\usepackage{amssymb}
\usepackage{graphicx}
\usepackage{epsfig}
\usepackage{epstopdf}
\usepackage{psfrag}
\usepackage{amsmath}
\usepackage{array}
\usepackage{subfigure}
\usepackage{cite,graphicx,amsmath,amssymb,color}
\usepackage{anysize}
\usepackage{algorithmic}
\usepackage{algorithm}

\usepackage{algorithm}
\usepackage{algorithmic}
\usepackage{stmaryrd}
\usepackage{multirow}
\usepackage{subfig}
\usepackage{graphicx,times,amsmath} 

\usepackage{url}

\marginsize{13mm}{13mm}{19mm}{43mm}

\begin{document}


\title{Enhanced VIP Algorithms for Forwarding, Caching, and Congestion Control in Named Data Networks}

\author{\authorblockN{Ying Cui, \ Fan Lai}\authorblockA{Shanghai Jiao Tong University, China}\and \authorblockN{Edmund Yeh, \ Ran Liu}\authorblockA{Northeastern University, USA}}

\maketitle

\newtheorem{Thm}{Theorem}
\newtheorem{Lem}{Lemma}
\newtheorem{Cor}{Corollary}
\newtheorem{Def}{Definition}
\newtheorem{Exam}{Example}
\newtheorem{Alg}{Algorithm}
\newtheorem{Sch}{Scheme}
\newtheorem{Prob}{Problem}
\newtheorem{Rem}{Remark}
\newtheorem{Proof}{Proof}
\newtheorem{Asump}{Assumption}
\newtheorem{Subp}{Subproblem}


\begin{abstract}
Emerging Information-Centric Networking (ICN) architectures seek to optimally utilize both bandwidth and storage for efficient content distribution over the network.
The Virtual Interest Packet (VIP) framework has been proposed to enable joint design of forwarding, caching, and congestion control strategies within the Named Data Networking (NDN) architecture.  While the existing VIP algorithms exhibit good performance, they are primarily focused on maximizing network throughput and utility, and do not explicitly consider user delay.  In this paper,  we develop a new class of enhanced algorithms for joint dynamic forwarding, caching and congestion control within the VIP framework.  These enhanced VIP algorithms adaptively stabilize the network and maximize network utility, while improving the delay performance by intelligently making use of VIP information beyond one hop.
Generalizing Lyapunov drift techniques, we prove the throughput optimality and characterize the utility-delay tradeoff of the enhanced VIP  algorithms.  Numerical experiments demonstrate the superior performance of the resulting enhanced algorithms for handling  Interest Packets and Data Packets  within the actual plane, in terms of low network delay and high network utility.

\end{abstract}





\section{Introduction}


It is increasingly clear that traditional connection-based networking architectures are ill suited for the prevailing user demands for network content~\cite{Zhang10}.
Emerging Information-Centric Networking (ICN) architectures aim to remedy this fundamental mismatch so as to dramatically improve the efficiency of content dissemination over the Internet.
In particular, Named Data Networking (NDN)~\cite{Zhang10}, or Content-Centric Networking (CCN)\cite{Jacobson},
is a proposed network
architecture for the Internet that replaces the traditional client-server  communication model   with
one based on the identity of data or content.

 Content delivery in NDN is accomplished using {\em Interest Packets} and {\em Data Packets}, along with specific data structures in nodes such as the {\em Forwarding Information Base (FIB)}, the {\em Pending Interest Table (PIT)}, and the {\em Content Store} (cache).  Communication is initiated by a data consumer or requester sending a request for the data using an \emph{Interest Packet}. Interest Packets are   forwarded along routes determined by the FIB at each node. Repeated requests for the same object can be suppressed at each node according to its PIT.
The {\em Data Packet} is subsequently transmitted back along the path taken by the corresponding Interest Packet, as recorded by the PIT at each node.
A node may optionally cache the data objects contained in the  received Data Packets in its local {\em Content Store}.
Consequently, a request for a data object can be fulfilled not only by the content source but also by any node with a copy of that object in its cache. Please see~\cite{Zhang10,VIPICN14,VIPfull16} for details.

NDN seeks to optimally utilize both bandwidth and storage for efficient content distribution, which highlights the need for joint design of traffic engineering and caching strategies, in order to optimize network performance.
To address this fundamental problem, in our previous work \cite{VIPICN14}, we propose the {\em VIP framework}  for the design of high performing NDN networks.
 Within this VIP framework,  we develop joint dynamic forwarding, caching and congestion control algorithms operating on virtual interest packets (VIPs)    to maximize network utility subject to network stability in the virtual plane, using Lyapunov drift techniques\cite{VIPICN14,VIPfull16}.   Then, using the resulting flow rates and queue lengths of the VIPs in the virtual plane, we develop the corresponding  joint dynamic  algorithms  in the actual plane, which have been shown to achieve superior performance in terms of network throughput, user utility, user delay, and cache hit rates, relative to several baseline policies.


While the VIP algorithms in \cite{VIPICN14,VIPfull16} exhibit excellent performance, they are primarily focused on maximizing network throughput and do not explicitly consider user delay.
 In this paper, we aim to further improve the delay performance of the existing VIP algorithms  in \cite{VIPICN14,VIPfull16} by leveraging VIP information beyond one hop.   There are several potential challenges in pursuing this. First, it is
not clear how one should improve the delay performance of
the existing VIP algorithms by jointly modifying forwarding, caching and congestion control
in a tractable manner. Second, it is not
clear how to maintain the desired throughput optimality and utility-delay tradeoff of
the existing VIP algorithms when the Lyapunov-drift-based control structure is modified
for improving the delay performance.

 In the following, we shall
address the above questions and challenges.
We first develop a new class of enhanced distributed forwarding and caching algorithms operating on VIPs to stabilize  network
in  the virtual plane.  We then extend the algorithm to include congestion control, thus achieving a favorable utility-delay tradeoff.
These enhanced VIP algorithms reduce  the delay  of the existing VIP algorithms by 1) exploiting the margin between the  VIP arrival rate vector   and the boundary of the VIP network stability region, and 2) making use of VIP information  beyond one hop in a simple and flexible manner.
Generalizing Lyapunov drift techniques, we demonstrate the throughput optimality and characterize the utility-delay tradeoff of the enhanced VIP algorithms.  These  enhanced VIP algorithms generalize the VIP  algorithms  in \cite{VIPICN14,VIPfull16} in the sense that they maintain network stability and maximize network utility  while improving delay performance. In addition, these enhanced VIP algorithms (designed for NDN networks) extend the enhanced dynamic backpressure algorithms  in \cite{CuiEnhancedBPTON15}  (designed for traditional source-destination networks)  in the sense that they incorporate caching into the joint design of  dynamic forwarding (routing) and congestion control.
Numerical experiments   demonstrate the superior performance of the resulting enhanced algorithms for handling  Interest Packets and Data Packets  within the actual plane, in terms of low network delay and high network utility.

Although there is now a rapidly growing literature in ICN, the problem of optimal joint forwarding and caching for content-oriented networks remains challenging.  In \cite{RossiniICN14}, the authors demonstrate the gains of joint forwarding and caching in ICNs. 
In \cite{PotentialRouting2012}, a potential-based forwarding scheme with random caching is proposed for ICNs.
The results in~\cite{PotentialRouting2012} are heuristic in the sense that it remains unknown how to choose proper potential values to  ensure good performance.  In \cite{CachingTM2011:Ying},   the authors propose throughput-optimal one-hop routing and caching to support the maximum number of requests in a single-hop Content Distribution Network (CDN) setting.
In \cite{TECC2012},  assuming the path between any two nodes is predetermined, the authors consider single-path routing (equivalently cache node selection) and caching to minimize link utilization for a general multi-hop content-oriented network.
The benefits of selective caching based on the concept of betweenness centrality, relative to ubiquitous caching, are shown in~\cite{Chai:2012:CLM:2342042.2342046}.
In~\cite{Age-based:6193504}, cooperative caching schemes  have been heuristically designed without being jointly optimized with forwarding strategies.  Finally, adaptive multi-path forwarding in NDN has been examined in~\cite{Yi:2012:AFN:2317307.2317319}, but has not been jointly optimized with caching strategies.

\section{Network Model}

\label{sec:model}

 We consider the same network model as in \cite{VIPICN14,VIPfull16}, which we describe for completeness.
{Consider a connected multi-hop (wireline) network modeled by a directed graph $\mathcal G=(\mathcal N, \mathcal L)$, where $\mathcal N$ and $\mathcal L$ denote the sets of $N$ nodes and $L$ directed links, respectively.  Assume  that $(b,a) \in {\cal L}$ whenever $(a,b) \in {\cal L}$.  Let $C_{ab} > 0$ be the transmission capacity (in bits/second) of link $(a,b) \in {\cal L}$.  Let $L_n\geq 0$ be the cache size (in bits) at node $n \in {\cal N}$.


Assume that content in the network are identified as {\em data objects},  each consisting of multiple data chunks.
Content delivery in NDN operates at the level of data chunks.  That is, each Interest Packet requests a particular data chunk, and a matching Data Packet consists of the requested data chunk, the data chunk name, and a signature.    A request for a data object consists of a sequence of Interest Packets which request all the data chunks of the object.
We consider a set ${\cal K}$ of $K$ data objects, which may be determined by the amount of control state that the network is able to maintain, and
 may include only the most popular data objects in the network, typically responsible for most of the network congestion\cite{VIPfull16}.
For simplicity, we assume that
all data objects have the same size $D$ (in bits).   The results in the paper can be
extended to the more general case where object sizes differ.   We consider
the scenario where  $L_n < KD$ for all $n \in
{\cal N}$.  Thus, no node can cache all data objects.
For each data object $k \in {\cal K}$, assume
that there is a unique node $src(k) \in {\cal N}$ which serves as the
content source for the object.    Interest Packets for chunks of a given data object
can enter the network at any node, and exit the network upon being satisfied
by matching Data Packets at the content source for the object, or at the
nodes which decide to cache the object.

\section{VIP Framework}

\label{sec:vipframe}

\begin{figure}[t]
\begin{center}
\includegraphics[width=70mm,height=20mm]{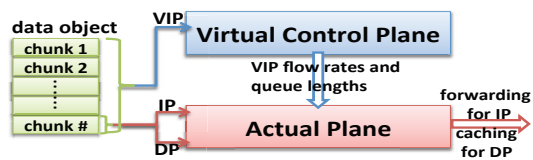}
\caption{\small{VIP framework \cite{VIPICN14}.  IP (DP) stands for Interest Packet (Data Packet).}}\label{fig:planes}
\end{center}
\end{figure}

We adopt the VIP framework proposed in \cite{VIPICN14}. In the following, we  briefly introduce the VIP framework to facilitate the discussion of
the algorithms developed in later sections. Please refer to \cite{VIPICN14} for the details on the motivation and utility of this framework.
As illustrated in Fig.~\ref{fig:planes}, the VIP framework  relies on virtual interest packets (VIPs), which
capture the {\em measured demand} for the respective data objects in the network, and represent content popularity which is empirically measured, rather than being given a priori.
The VIP framework employs a \emph{virtual} control plane  operating on VIPs {\em at the data object level}, and an \emph{actual} plane  handling Interest Packets and Data Packets {\em at the data chunk level}.
 The virtual plane facilitates the design of  distributed control algorithms operating on VIPs, aimed at yielding desirable performance in terms of network metrics of concern,  by taking advantage of local information on network demand (as represented by the VIP counts).
The flow rates and queue lengths of the VIPs resulting from the control algorithm in the virtual plane are then used to specify the control algorithms in the actual plane \cite{VIPICN14}.



We now specify the dynamics of the VIPs within the virtual plane.
Consider time slots of length 1 (without loss of generality) indexed by $t = 1, 2, \ldots$.  Specifically, time slot $t$ refers to the time interval $[t,t+1)$.
Within the virtual plane, each node $n\in {\cal N}$ maintains a separate VIP queue  for each data object $k \in {\cal K}$.
Note that no data is contained in these VIPs.  Thus, the VIP queue size for each node $n$ and data object $k$ at the beginning of slot $t$   is  represented by a {\em counter} $V_n^k(t)$.\footnote{We assume that VIPs can be quantified as a real number.  This is reasonable when the VIP counts are large.}
An exogenous request for data object $k$ is considered to have arrived at node $n$ if the Interest Packet requesting the starting chunk of data object $k$ has arrived at node $n$.
Let $A^k_n(t)$ be the number of exogenous data object request arrivals at node $n$ for object $k$ during slot $t$.\footnote{We think of a node  as an aggregation  point  combining many network users, 
and hence it is likely to submit many requests for a data object over time.}
For every arriving request for data object $k$ at node $n$, a corresponding VIP for object $k$ is generated at $n$.
The long-term exogenous VIP arrival rate at node $n$ for object $k$ is
$ \lambda_n^k\triangleq \mathbb \limsup_{t \rightarrow \infty} \frac{1}{t} \sum_{\tau = 1}^t A^k_n(\tau).$
Let $\mu_{ab}^k(t)  \geq 0$ be the allocated transmission rate of VIPs for data object $k$ over link $(a,b)$ during time slot $t$.  Note that
a single message between node $a$ and node $b$ can summarize all the VIP transmissions during  each slot. 
Data Packets for the requested data object must travel on the reverse path taken by the Interest Packets.  Thus, in determining the transmission of the VIPs, we  consider the link capacities on the reverse path as follows:
\begin{align}
&\sum_{k \in \mathcal K} {\mu^{k}_{ab}(t)} \leq C_{ba}/D,\
\text{for~all}~(a,b) \in \mathcal L\label{eqn:rout_cost_sum}\\
&\mu^{k}_{ab}(t)=0, \;\text{for~all}~(a,b)\not\in \mathcal
L^{k}\label{eqn:rout_cost_non_neg}
\end{align}
where $C_{ba}$ is the capacity of ``reverse" link $(b,a)$ and  $\mathcal L^k$ is the set of $L^k$ links which are allowed to transmit the VIPs of object $k$.  Let $C_{\max}\triangleq \max_{(a,b)\in\mathcal L}C_{ab}/D$.

In the virtual plane, we may assume that at each slot $t$, each node $n \in {\cal N}$ can gain access to any data object $k \in {\cal K}$ for which there is interest at $n$, and potentially cache the object locally.
Let $s_n^{ k}(t) \in \{0,1\}$ represent the caching state for object $k$ at node $n$ during slot $t$, where $s_n^{k}(t)=1$ if object $k$  is cached at node $n$ during slot $t$, and $s_n^{k}(t)=0$ otherwise.  Note that even if $s_n^{k}(t)=1$, the content store at node $n$ can satisfy only a limited number of VIPs during one time slot.  This is because there is a maximum rate $r_n$ (in objects per slot) at which node $n$ can produce copies of cached object $k$ \cite{VIPICN14,VIPfull16}.


Initially, all VIP counters are set to 0, i.e., $V_n^k(1)=0$.
The time evolution of the VIP count at node $n$ for object $k$ is as follows:
\begin{small}
\begin{align}
& V^k_n(t+1) \leq \nonumber\\
&   \left(
\left(V^k_n(t)-\sum_{b\in \mathcal N}\mu^{k}_{nb}(t)\right)^+ +A^k_n(t)
+\sum_{a\in \mathcal N}\mu^{k}_{an}(t)- r_n s_n^{k}(t)\right)^+
\label{eqn:queue_dyn}
\end{align}
\end{small}
where $(x)^+ \triangleq \max(x,0)$.  Furthermore,
$V^k_n(t) = 0$ for all $t \geq 1$ if $n=src(k)$.  The detailed explanation of \eqref{eqn:queue_dyn} can be found in \cite{VIPICN14,VIPfull16}. Physically, the VIP count can be interpreted as a {\em potential}.  For any  data object, there is a downward ``gradient" from entry points of the data object requests to the content source and caching nodes.




The {\em VIP network stability region} $\Lambda$ is the closure
of the set of all VIP arrival rates $\boldsymbol \lambda\triangleq (\lambda^k_n)_{n \in {\cal N},k \in {\cal K}}$ for which there exists some feasible (i.e., satisfying~\eqref{eqn:rout_cost_sum}-\eqref{eqn:rout_cost_non_neg} and the cache size limits $(L_n)_{n \in {\cal N}}$)  joint forwarding and caching policy which can guarantee that all VIP queues are stable \cite{VIPICN14}.
Assume (i) the VIP arrival processes $\{A^k_n(t): t=1,2,\ldots\}$  are mutually independent with respect to $n$ and $k$; (ii)
for all $n $ and $k  $, $\{A^k_n(t): t = 1, 2, \ldots\}$ are i.i.d. with respect to $t$; and (iii) for all $n$ and $k$, $A^k_n(t)\leq A^k_{n,\max}$ for all $t$.  Under these assumptions, Theorem~1 in \cite{VIPfull16} characterizes the VIP stability region in the virtual plane (or equivalently the Interest Packet stability region in the actual plane when there is no collapsing or suppression at the PITs).  Note that the theoretical results in this paper also hold under these assumptions.

In the following, with the aim of improving the delay performance of the VIP algorithms in \cite{VIPICN14,VIPfull16}, we focus on developing  a new class of enhanced VIP algorithms within the virtual plane of the VIP framework, for the cases where  $\boldsymbol \lambda  \in  \text{int}(\Lambda)$ and $\boldsymbol \lambda \notin \Lambda$ in Sections~\ref{sec:forwarding-caching-VIP} and \ref{sec:congestion-VIP}, respectively.

\section{Enhanced Throughput Optimal VIP Control}\label{sec:forwarding-caching-VIP}


In this section, we consider the case where $\boldsymbol \lambda\in \text{int}(\Lambda)$, and develop a new class of enhanced  joint dynamic forwarding and caching algorithms, within the virtual plane of the VIP framework.

%

\subsection{Bias Function}

The VIP algorithm, i.e., Algorithm~1 in \cite{VIPICN14,VIPfull16}, focuses primarily on maximizing network throughput, and uses one-hop VIP count differences for forwarding and on per-node VIP counts for caching.  This leads to a simple distributed implementation.  On the other hand, by incorporating VIP count information beyond one hop in a tractable manner, one can potentially improve the delay performance of the VIP algorithm while retaining the desirable throughput optimality and distributed implementation.
Toward this end,  we consider a general nonnegative VIP count-dependent  bias function for each node $n\in \mathcal N$ and object $k\in \mathcal K$  \cite{CuiEnhancedBPTON15}:
\begin{align}
f_n^{k}(\mathbf
v)=\sum_{n'\in \mathcal N}\eta^{k}_{nn'}(\mathbf
v) v^{k}_{n'}/z_{n'}^{k}. \label{eqn:bias-func}
\end{align}
Here,  $\mathbf v\triangleq (v^k_n)_{n \in \mathcal N,k\in \mathcal K}$ represents the VIP counts at a particular time slot  and $\eta^{k}_{nn'}(\mathbf
v)\in [0,1]$ is the weight associated with VIP count $v^{k}_{n'}$ at node $n$ for object $k$, representing the relative importance of $v^{k}_{n'}$ in the bias at node $n$ for object $k$.  The parameter $z_{n'}^{k}>0$ is designed to
guarantee network stability and  will be discussed below in Theorem~\ref{Thm:thpt-opt} and Theorem~\ref{Thm:flow-control}.
We can treat $v^{k}_{n'}/z_{n'}^{k}$  as a normalized version of $v^{k}_{n'}$.
While the bias function  in \eqref{eqn:bias-func}  is generally written as a function of the global  VIP counts,
one can choose the bias function to depend only on the local VIP counts within one hop  \cite{CuiEnhancedBPTON15}. For example, as in \cite{CuiEnhancedBPTON15},  we can choose the  minimum next-hop VIP count bias function, i.e.,
\begin{align}
f_n^{k}\left(\mathbf v_n^{k}\right)=\frac{1}{ z}\min_{n'\in \left\{n': (n,n')\in \mathcal L^{k}\right\}} v_{n'}^{k}\label{eqn:ex-onehop-f}
\end{align}
where  $\mathbf v_n^k\triangleq (v^k_{n'})_{n'\in \left\{n': (n,n')\in \mathcal L^{k}\right\}}$.


%

Each specific choice of a bias function $\mathbf f\triangleq (f_n^{k})_{n\in\mathcal N, k\in\mathcal K}$ corresponds to one enhanced  VIP algorithm, and the number of VIP counts contributing to the bias function determines the implementation complexity of the corresponding enhanced VIP algorithm.  The form of the bias function is carefully chosen to stabilize the network or maximize the network utility, while at the same time offering a high degree of flexibility in choosing specific enhanced VIP algorithms with manageable complexity, distributed implementation, and significantly better delay performance\cite{CuiEnhancedBPTON15}.

\subsection{Enhanced Forwarding and Caching Algorithm}

 We now present a new class of enhanced   joint dynamic  forwarding and caching algorithms for VIPs in the virtual plane by incorporating the general  VIP count-dependent bias function in \eqref{eqn:bias-func} into Algorithm~1 in \cite{VIPICN14,VIPfull16}.
\begin{Alg}[Enhanced Forwarding and Caching]  At the beginning of each slot  $t$, observe the VIP counts $\mathbf V(t)\triangleq (V^k_n(t))_{ n \in \mathcal N,k\in \mathcal K}$ and perform forwarding and caching in the virtual plane as follows.

\textbf{Forwarding}: For each data object $k \in {\cal K}$ and each link $(a,b)\in \mathcal
L^k$,  choose
\begin{align}
\mu^{k}_{ab}(t)
=&\begin{cases} C_{ba}/D,
&  W^*_{ab}(t)>0\  \text{and}\ k=k^*_{ab}(t)\\
0, &\text{otherwise}
\end{cases}\label{eqn:forwarding-VIP}
\end{align}
where $W_{ab}^{k}(t) \triangleq \left(V^k_a(t)+f_a^k(\mathbf V(t))\right) - \left(V^k_b(t)+f_b^k(\mathbf V(t))\right)$, $k^*_{ab}(t) \triangleq \arg
\max_{k\in\{k: (a,b)\in \mathcal L^{k}\}} W_{ab}^{k}(t)$, and $W^*_{ab}(t)
\triangleq \left(W_{ab}^{k^*_{ab}(t)}(t)\right)^+$.
Here,
$W_{ab}^{k}(t)$ is the  enhanced  backpressure weight of object $k$ on link $(a,b)$ at slot $t$, and
$k^*_{ab}(t)$ is the data object which maximizes the  enhanced  backpressure weight on link $(a,b)$ at time $t$.

\textbf{Caching}:  At each node $n \in \mathcal N$, choose $(s^k_n(t))_{n\in\mathcal N, k\in\mathcal K}$ to
\begin{equation}
\max \sum_{k\in \mathcal K}\left(V^k_n(t)+f_n^k(\mathbf V(t)\right) s^k_n \quad
\text{s.t.} \ \sum_{k\in \mathcal K} s^k_n\leq L_n/D.
\label{eqn:knapsack-VIP}
\end{equation}
 Here, $V^k_n(t)+f_n^k(\mathbf V(t))$ represents the enhanced caching weight of object $k$ at node $n$.

Based on the forwarding and caching in \eqref{eqn:forwarding-VIP} and \eqref{eqn:knapsack-VIP}, the VIP count is updated according to \eqref{eqn:queue_dyn}.  For simplicity, we use the same  bias function for  forwarding and caching. In general, the bias functions for forwarding and caching can be different.
\label{Alg:VIP}
\end{Alg}

\begin{Rem} [Connection to the  Algorithms in  \cite{CuiEnhancedBPTON15} and \cite{VIPICN14,VIPfull16}] Algorithm~\ref{Alg:VIP}  (designed for NDN networks) extends the enhanced dynamic backpressure algorithms  in \cite{CuiEnhancedBPTON15}  (designed for traditional source-destination networks)  in the sense that it incorporates caching into the joint design of  dynamic forwarding and congestion control.
Algorithm~\ref{Alg:VIP} also generalizes Algorithm~1 in \cite{VIPICN14,VIPfull16} in the sense that it maintains network stability while improving delay performance by incorporating VIP information beyond one hop.\end{Rem}


At each slot $t$ and for each link $(a,b)$, the enhanced backpressure-based forwarding algorithm allocates the entire normalized ``reverse" link capacity $C_{ba}/D$ to transmit the VIPs for the data object $k^*_{ab}(t)$ which maximizes the enhanced  backpressure $W_{ab}^{k}(t) $.  
 The enhanced  max-weight  caching algorithm  implements the optimal solution to the   max-weight knapsack problem in \eqref{eqn:knapsack-VIP}, i.e., allocate cache space at node $n$ to the $\lfloor L_n/D\rfloor$ objects with the largest enhanced caching weights $V^k_n(t)+f_n^k(\mathbf V(t))$.
The enhanced forwarding and caching algorithm maximally balances out the VIP counts by joint forwarding and caching, to prevent
congestion building up in any part of the network, thereby reducing delay.



It is important to note that with local VIP count-bias functions, such as the minimum next-hop VIP count bias function in \eqref{eqn:ex-onehop-f}, both the enhanced  backpressure-based forwarding algorithm and the enhanced  max-weight caching algorithm are {\em distributed}. Following the complexity analysis  in \cite{CuiEnhancedBPTON15}, we know that the enhanced VIP algorithm in Algorithm~\ref{Alg:VIP} with  the minimum next-hop VIP bias function in \eqref{eqn:ex-onehop-f}  has the same order of implementation complexity as Algorithm~1   in \cite{VIPICN14,VIPfull16} (without any bias functions).  In general, for ease of implementation, one should intelligently choose the VIP counts which contribute to the bias function, leading to enhanced VIP  algorithms with distributed implementation and good delay performance.}

\subsection{Throughput Optimality}


We now show that   Algorithm~\ref{Alg:VIP} adaptively stabilizes all VIP queues for any ${\boldsymbol \lambda} \in {\rm
int}({\Lambda})$, without knowing ${\boldsymbol \lambda}$.

\begin{Thm} [Generalized Throughput Optimality] Given  $\boldsymbol \epsilon\triangleq(\epsilon_n^k)_{n \in {\cal N}, k \in {\cal K}} \succeq \mathbf 0$
such
that $\boldsymbol \lambda+\boldsymbol \epsilon \in  \text{int}(\Lambda)$, there exist $\boldsymbol \delta\triangleq(\delta_n^{k})_{n \in {\cal N}, k \in {\cal K}}  \succ \mathbf 0$ such that $\boldsymbol \lambda+\boldsymbol \epsilon+\boldsymbol \delta \in \Lambda$,
and $\mathbf z\triangleq(z{_n^{k}})_{n \in {\cal N}, k \in {\cal K}}  \succ \mathbf 0$ such that $\boldsymbol \epsilon_z\triangleq\bigg(\dfrac{2C_{\max}L^{k}+N r_{\max}}{z_n^{k}}\bigg)_{n \in {\cal N}, k \in {\cal K}}  \preceq \boldsymbol \epsilon$. Then, the network of VIP queues under
Algorithm \ref{Alg:VIP} satisfies
\begin{small}
\begin{align}
\limsup_{t\to\infty}\frac{1}{t}\sum_{\tau=1}^{t}\sum_{n\in \mathcal N,k\in \mathcal K} \mathbb
E[V^k_n(\tau)]\leq \frac{N B}{\beta_z}\label{eqn:enhanced-DBP-VB}
\end{align}\end{small}
where
$ B\triangleq \frac{1}{2N}\sum_{n\in \mathcal N}\big((\mu^{out}_{n,
\max})^2+(A_{n,\max}+\mu^{in}_{n,\max}+Kr_n)^2
+2\mu^{out}_{n,
\max}Kr_n\big)$ ,
$\beta_z=\sup_{\substack{\{(\boldsymbol \epsilon,\boldsymbol \delta): \boldsymbol \epsilon \succ \boldsymbol \epsilon_z,\boldsymbol \delta \succ \mathbf 0,\\ \boldsymbol \lambda+\boldsymbol \epsilon+\boldsymbol \delta \in \in \Lambda\}}} \min\limits_{n\in \mathcal N,k\in \mathcal K}\bigg\{\epsilon_n^{k}+\delta_n^{k}-\dfrac{2C_{\max}L^{k}+Nr_{\max}}{z_n^k}\bigg\},$
with
$\mu^{in}_{n, \max}\triangleq \sum_{a\in \mathcal N} C_{an}/D $, $\mu^{out}_{n, \max}\triangleq
\sum_{b\in \mathcal N} C_{nb}/D$, $
A_{n,\max} \triangleq \sum_{k\in \mathcal K}
A^k_{n,\max}$, and $r_{\max}\triangleq \max_{n\in\mathcal N} r_n$.\label{Thm:thpt-opt}
\end{Thm}

Similar to Theorem~1 in \cite{CuiEnhancedBPTON15}, Theorem~\ref{Thm:thpt-opt} should be interpreted as follows.  When it is given that  $\boldsymbol \lambda$ is bounded away from the boundary of   $\Lambda$ by at least $\boldsymbol \epsilon \succ \mathbf 0$, i.e., $\boldsymbol \lambda+\boldsymbol \epsilon \in \text{int}(\Lambda)$,
one can choose a {\em finite} $\mathbf z \succ \mathbf 0$ such that $\boldsymbol \epsilon_{\mathbf z}\prec \boldsymbol \epsilon$.
In this case, Algorithm~\ref{Alg:VIP}  can improve the delay performance of  Algorithm~1 in \cite{VIPICN14,VIPfull16} (which will be demonstrated numerically in   Section~\ref{sec:simulations}) while maintaining a generalized notion of throughput optimality, by exploiting the margin $\boldsymbol \epsilon \succ \mathbf 0$ to incorporate
VIP counts beyond one-hop \cite{CuiEnhancedBPTON15}.
When it is only known that  $\boldsymbol \lambda \in \text{int}(\Lambda)$ and no extra margin is given ($\boldsymbol \epsilon = \mathbf 0$), then by Theorem~\ref{Thm:thpt-opt}, $z_n^{k}$  must be  chosen to be infinity for all $n\in \mathcal N$ and $k\in \mathcal K$ (i.e., $\mathbf f = \mathbf 0$).  In this case, Algorithm~\ref{Alg:VIP} reduces to Algorithm~1 in \cite{VIPICN14,VIPfull16}, and Theorem~\ref{Thm:thpt-opt} reduces to Theorem~2 in \cite{VIPfull16}. Theorem~\ref{Thm:thpt-opt} can be seen as the generalization of the throughput optimal results in Theorem~1 of \cite{CuiEnhancedBPTON15} and  Theorem~2 of \cite{VIPfull16}.

\section{Enhanced VIP Congestion Control}\label{sec:congestion-VIP}

The VIP forwarding and caching algorithm first described in~\cite{VIPICN14} were extended to incorporate congestion control in~\cite{VIPfull16}.  Here, we develop a new class of enhanced algorithms which generalize the enhanced forwarding and caching algorithm (Algorithm 1) described above to incorporate congestion control.


\subsection{Transport Layer and Network Layer VIP Dynamics}

Even with throughput optimal forwarding and caching, excessively large request rates ($\boldsymbol \lambda\not\in \Lambda$) can overwhelm a NDN network with limited resources. When $\boldsymbol \lambda\not\in \Lambda$,  newly arriving Interest Packets (equivalently VIPs) first enter transport-layer storage reservoirs before being admitted to network-layer queues.
Let $Q_{n,\max}^k$ and $Q_n^k(t)$ denote  the transport layer VIP buffer size and VIP count for object $k$ at node $n$ at the beginning of slot $t$, respectively.
Let $\alpha^k_n(t)\geq 0$ denote the amount of VIPs admitted to the network layer VIP queue of object $k$ at node $n$ from the transport layer VIP queue at slot $t$. Assume $\alpha^k_n(t)\leq \alpha^k_{n,\max}$, where $\alpha^k_{n,\max}$ is a   positive constant which limits the burstiness of the admitted VIPs to the network layer.  We have  the following time evolutions of the transport  and network layer  VIP counts\cite{Georgiadis-Neely-Tassiulas:2006,VIPfull16}:

\begin{small}\begin{align}
&Q^k_n(t+1) = \min \left\{ \left(Q^k_n(t)-\alpha^k_n(t)\right)^+ +A^k_n(t), Q^k_{n,\max} \right\}
\label{eqn:queue_dyn-trans}\\
& V^k_n(t+1) \leq \nonumber\\
&   \left(
\left(V^k_n(t)-\sum_{b\in \mathcal N}\mu^{k}_{nb}(t)\right)^+ +\alpha^k_n(t)
+\sum_{a\in \mathcal N}\mu^{k}_{an}(t)- r_n s_n^{k}(t)\right)^+.\label{eqn:queue_dyn-flow}
\end{align}
\end{small}

\subsection{Enhanced Congestion Control Algorithm}

The goal of congestion control is to support a portion of the VIPs  which maximize the sum utility when $\boldsymbol \lambda  \notin \Lambda$.  Let $g^k_n(\cdot)$ be the utility function associated with the VIPs admitted into the network layer for object $k$ at node $n$.  Assume  $g^k_n(\cdot)$ is non-decreasing, concave, continuously differentiable and non-negative. Define a $\boldsymbol \theta$-optimal admitted VIP rate\cite{Georgiadis-Neely-Tassiulas:2006,VIPfull16}:
\begin{align}
\overline {\boldsymbol \alpha}^*(\boldsymbol \theta)\triangleq\arg\max_{\overline{\boldsymbol \alpha}}\quad  & \sum_{n\in \mathcal N,k\in \mathcal K}g^k_n\left(\overline \alpha^k_n\right)\label{eqn:eps-opt-prob}\\
s.t. \quad &  \overline{\boldsymbol \alpha}+\boldsymbol \theta \in \Lambda
,\  \mathbf 0 \preceq \overline{\boldsymbol \alpha} \preceq \boldsymbol \lambda
\end{align}
where $ \overline{\boldsymbol \alpha}^*(\boldsymbol \theta)\triangleq(\overline \alpha^{k*}_n(\boldsymbol \theta))_{n \in \mathcal N,k\in \mathcal K}$, $ \overline{\boldsymbol \alpha}\triangleq(\overline \alpha^k_n)_{n \in \mathcal N,k\in \mathcal K}$ and $\mathbf 0\preceq\boldsymbol \theta\triangleq(\theta^k_n)_{n \in \mathcal N,k\in \mathcal K}\in \Lambda $.
Due to the non-decreasing property of the utility functions, the maximum sum utility over all $\boldsymbol \theta$ is achieved at $\overline {\boldsymbol \alpha}^*(\mathbf 0)$ when  $\boldsymbol \theta=\mathbf 0$.

We now develop a  new class of enhanced  joint congestion control, forwarding and caching algorithms that yield a throughput vector which can be arbitrarily close to the optimal solution $\overline {\boldsymbol \alpha}^*(\mathbf 0)$. We introduce auxiliary variables $\gamma_n^k(t)$ and the virtual VIP count $Y_n^k(t)$ for all $n\in \mathcal N$ and $k\in \mathcal K$.  Set $Y^k_n(0)=0$ for all $n\in\mathcal N$ and $k\in \mathcal K$.

\begin{Alg} {\em (Enhanced Congestion Control, Forwarding and Caching)}
At the beginning of  each  slot $t$,   observe the VIP counts $\mathbf V(t)$ and  $\mathbf Y(t)\triangleq (Y^k_n(t))_{n \in \mathcal N,k\in \mathcal K}$, and perform the following congestion control, forwarding
and caching in the virtual plane as follows.\footnote{Note that the congestion control part  of Algorithm \ref{Alg:flow-DBP} is the same as that in \cite[page 90]{Georgiadis-Neely-Tassiulas:2006} and  Algorithm~3 in \cite{VIPfull16}. The difference lies in the forwarding  and caching part. We describe the congestion control part here for the purpose of completeness.}

\textbf{Congestion Control}: For each node $n$ and object $k$, choose the admitted VIP count at slot $t$, which also serves as the output rate of the corresponding virtual queue:
\begin{align}
\alpha^k_n(t)=
\begin{cases}
\min\left\{Q^k_n(t), \alpha^k_{n,\max}\right\},  & Y^k_n(t)>V^k_n(t)\\
0,& \text{otherwise}.
\end{cases}
\end{align}
Then, choose the auxiliary variable, which serves as the input rate  to the corresponding virtual queue:
\begin{align}
\gamma^k_n(t)=\arg\max_{\gamma}\quad & W g^k_n(\gamma)-Y^k_n(t)\gamma\label{eqn:solu-gamma}\\
s.t. \quad & 0\leq \gamma\leq \alpha^k_{n,\max}\nonumber
\end{align}
where $W>0$ is a control parameter which affects the utility-delay tradeoff of the algorithm.
Based on the chosen $\alpha^k_n(t)$ and $\gamma^k_n(t)$, the  transport
layer VIP count is updated according to \eqref{eqn:queue_dyn-trans} and the virtual VIP count is  updated according to:
\begin{align}
Y^k_n(t+1)=& \left(Y^k_n(t)-\alpha^k_n(t)\right)^+ +\gamma^k_n(t).\label{eqn:queue_dyn-virtual}
\end{align}

\textbf{Forwarding and Caching}:  Same as Algorithm \ref{Alg:VIP}. The network layer VIP count is updated according to \eqref{eqn:queue_dyn-flow}.\label{Alg:flow-DBP}
\end{Alg}

\subsection{Utility-Delay Tradeoff}

 We now show that  by tuning  control parameter $W>0$,  Algorithm~\ref{Alg:flow-DBP}
adaptively achieves a utility-delay tradeoff for VIP queues, for any ${\boldsymbol \lambda} \not\in \Lambda$, without knowing ${\boldsymbol \lambda}$.  In addition, Algorithm~\ref{Alg:flow-DBP} yields a throughput vector which can be arbitrarily close to   $\overline {\boldsymbol \alpha}^*(\mathbf 0)$ by letting $W\to 0$.


\begin{Thm}[Utility-Delay Tradeoff] For an arbitrary VIP arrival rate  $\boldsymbol \lambda$ and for any control parameter $W>0$, given  $\boldsymbol \epsilon  \in  \text{int}(\Lambda)$, there exist $\boldsymbol \delta \succ \mathbf 0$ such that $\boldsymbol \epsilon+\boldsymbol \delta \in \Lambda$,
and $\mathbf z \succ \mathbf 0$ such that $\boldsymbol \epsilon_z \preceq \boldsymbol \epsilon$. Then,  the  network of VIP queues under
Algorithm \ref{Alg:flow-DBP}  satisfies
\begin{small}
\begin{align}
&\limsup_{t\to\infty}\frac{1}{t}\sum_{\tau=1}^{t}\sum_{n\in \mathcal N,k\in \mathcal K} \mathbb
E[V^k_n(\tau)]\leq \frac{2N\hat B+WG_{\max}}{2\hat \beta_z}
\label{eqn:enhanced-flow-DBP-U}\\
&\liminf_{t\to\infty}\sum_{n\in \mathcal N,k\in \mathcal K} g^k_n\left(\overline \alpha^k_n(t)\right)\geq
\sum_{n\in \mathcal N,k\in \mathcal K}
g^{k}_n\left( \overline \alpha^{k*}_n\left(\boldsymbol \epsilon_z\right)\right)-\frac{2N\hat B}{W}
\label{eqn:enhanced-flow-DBP-g}
\end{align}
\end{small}
where $\hat B\triangleq \frac{1}{2N}\sum_{n\in \mathcal N}\Big((\mu^{out}_{n,
\max})^2+(\alpha_{n,\max}+\mu^{in}_{n,\max}+Kr_n)^2+2\left(\alpha_{n,\max} \right)^2
+2\mu^{out}_{n,\max}Kr_n\Big)$ and $\hat \beta_z=
\sup_{\substack{\{(\boldsymbol \epsilon,\boldsymbol \delta):  \boldsymbol \epsilon\succeq \boldsymbol \epsilon_{\mathbf z}, \boldsymbol \delta \succ \mathbf 0, \\  \quad \ \boldsymbol \epsilon+\boldsymbol \delta\in  \Lambda\}}} \min\limits_{n\in \mathcal N,k\in \mathcal K}\bigg\{\epsilon_n^{k}+\delta_n^{k}
-\dfrac{2C_{\max}L^{k}+Nr_{\max}}{z_n^k}\bigg\}$
with $\alpha_{n,\max} \triangleq  \sum_{k\in \mathcal K}
\alpha^k_{n,\max}$,   $G_{\max}\triangleq\sum_{n\in \mathcal N,k\in \mathcal K}
g^k_n\left(\alpha^k_{n,\max}\right)$, $\overline \alpha^k_n(t)\triangleq \frac{1}{t}\sum_{\tau=1}^t\mathbb E[ \alpha^k_n(\tau)]$.
\label{Thm:flow-control}
\end{Thm}

Similar to Theorem~2 in \cite{CuiEnhancedBPTON15}, Theorem~\ref{Thm:flow-control} should be interpreted as
follows. When $\mathbf 0 \prec \boldsymbol \epsilon \in
\text{int}(\Lambda)$, one can choose a {\em finite} $\mathbf
z \succ \mathbf 0$ such that $\boldsymbol \epsilon_{\mathbf
z} \preceq
\boldsymbol \epsilon$.
In this case, Algorithm~\ref{Alg:flow-DBP}  can improve the utility-delay tradeoff of  Algorithm~3 in \cite{VIPfull16} (which will be demonstrated numerically in Section~\ref{sec:simulations}), by exploiting the margin $\boldsymbol \epsilon$ to incorporate  VIP counts beyond one-hop \cite{CuiEnhancedBPTON15}.
When $\boldsymbol \epsilon = \mathbf 0$, i.e., no margin is given,
$z_n^{k}$ is chosen to be infinity for all $n\in \mathcal N$ and $k\in \mathcal K$ (i.e., $\mathbf f = \mathbf 0$). In this case, Algorithm~\ref{Alg:flow-DBP}  reduces to
Algorithm~3 in \cite{VIPfull16}, and
Theorem~\ref{Thm:flow-control} reduces to Theorem~3 in \cite{VIPfull16}. Theorem~\ref{Thm:flow-control} can be seen as the generalization of the utility-delay tradeoff  results in Theorem~2 of \cite{CuiEnhancedBPTON15} and  Theorem~3 of \cite{VIPfull16}.

\section{Experimental Evaluation}
\label{sec:simulations}

\begin{figure}[t!]
\begin{center}
  \subfigure[\small{GEANT   (22 nodes).}]
  {\resizebox{4cm}{3.1cm}{\includegraphics{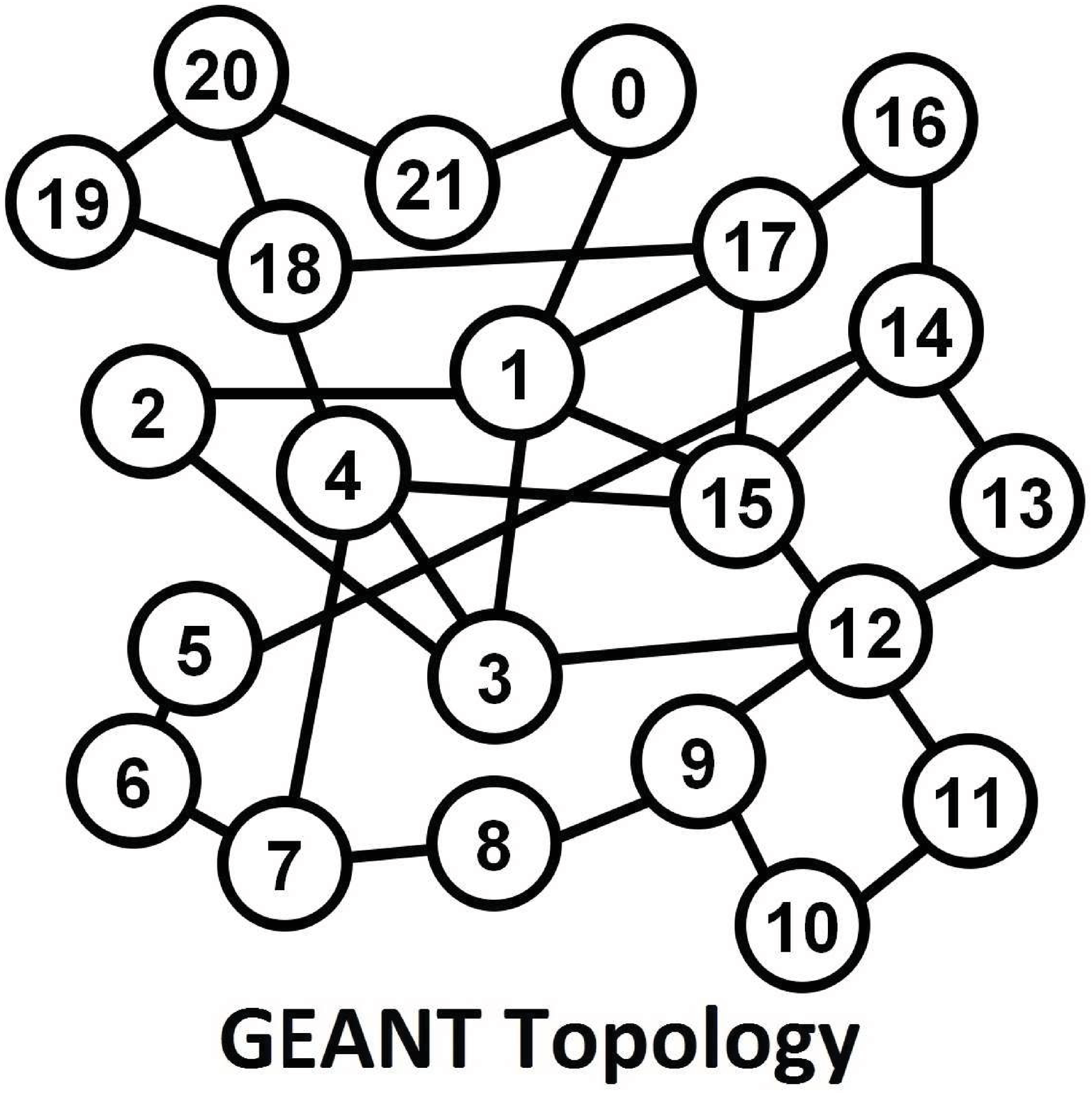}}}\
    \subfigure[\small{DTelekom   (68 nodes).}]
  {\resizebox{4cm}{3.1cm}{\includegraphics{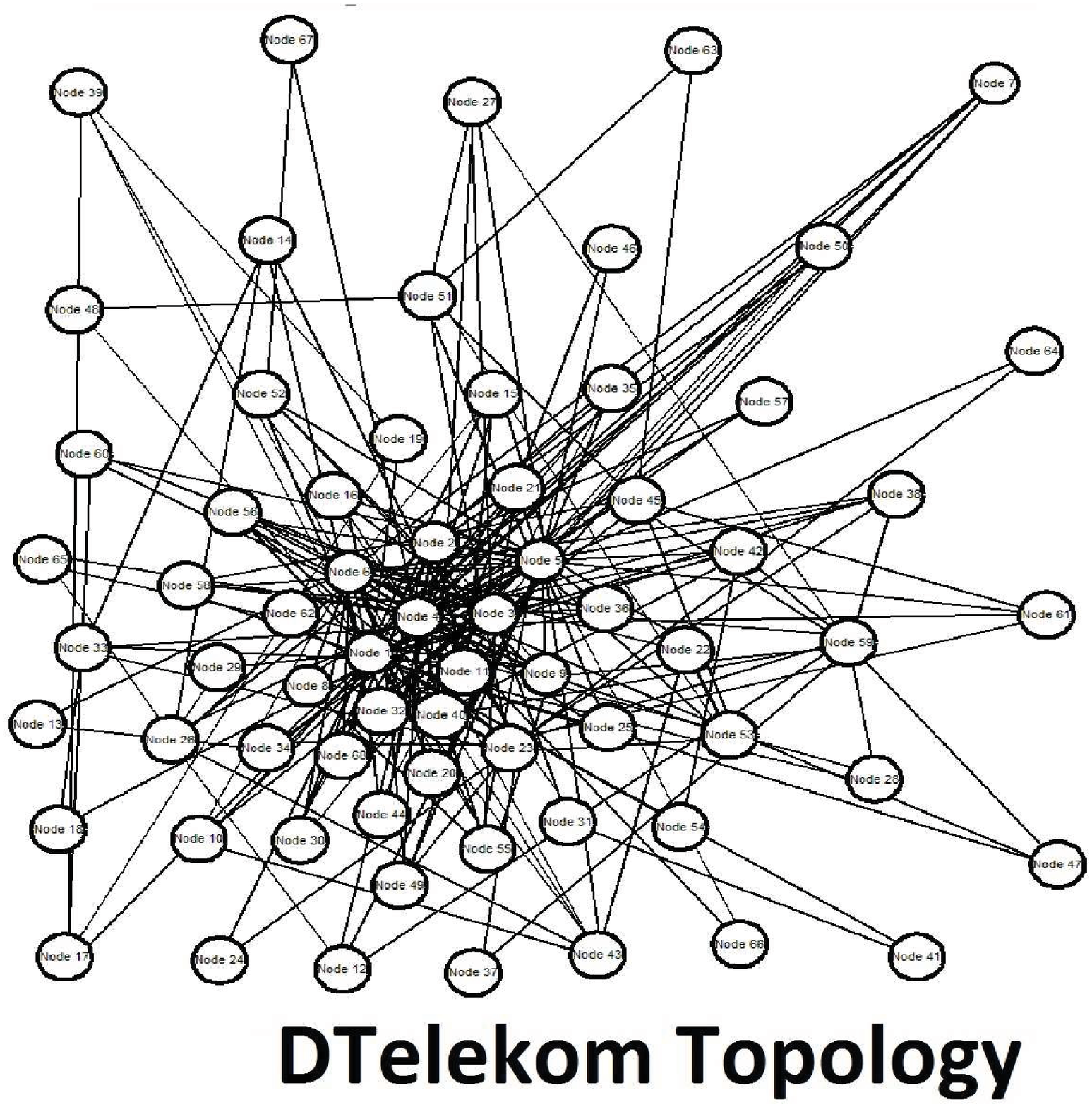}}}
  \end{center}
    \caption{\small{Network topologies.}}
\label{Fig:topo}
\end{figure}

\begin{figure}[t!]
\begin{center}
  \subfigure[\small{Delay for GEANT.}]
  {\resizebox{8.2cm}{!}{\includegraphics{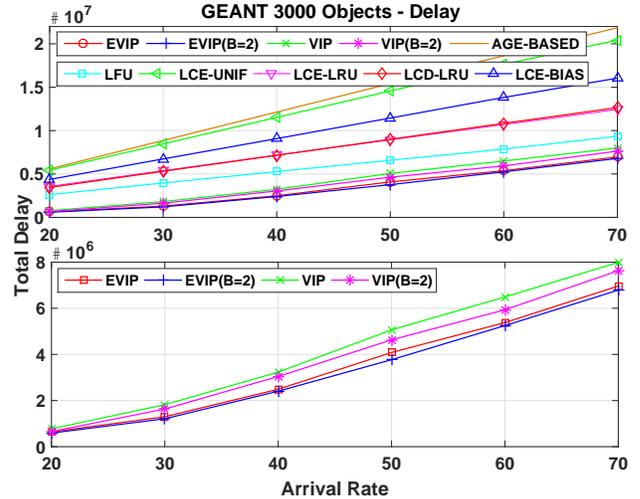}}}
    \subfigure[\small{Delay for DTelekom.}]
  {\resizebox{8.2cm}{!}{\includegraphics{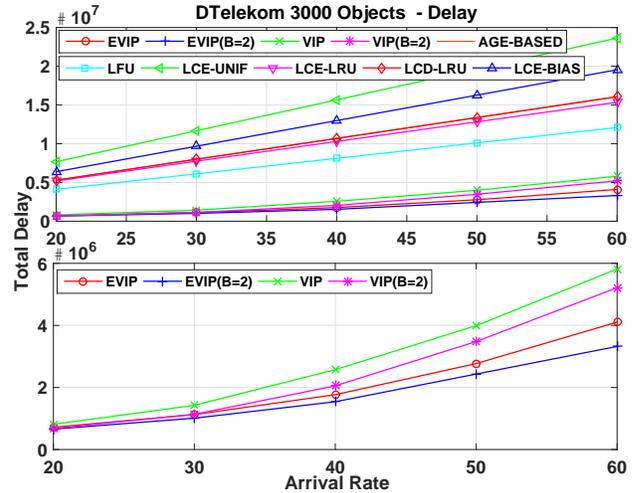}}}
  \end{center}
    \caption{\small{Average delay.}}
\label{Fig:delay}
\end{figure}

\begin{figure}[t!]
\begin{center}
  \subfigure[\small{Utility-delay tradeoff for GEANT.}]
  {\resizebox{8.2cm}{!}{\includegraphics{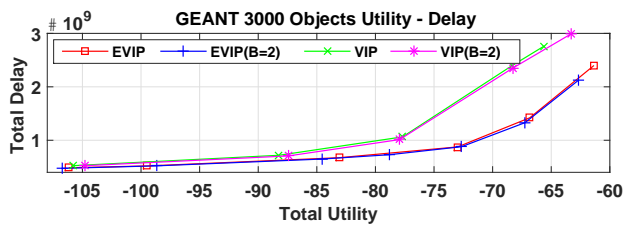}}}
    \subfigure[\small{Utility-delay tradeoff for DTelekom.}]
  {\resizebox{8.2cm}{!}{\includegraphics{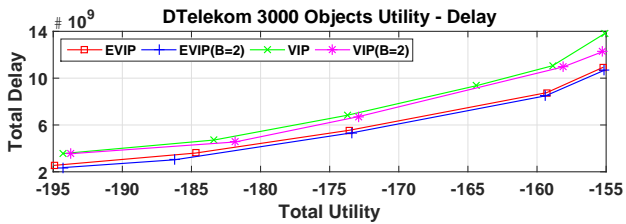}}}
  \end{center}
    \caption{\small{Utility-delay tradeoff.}}
\label{Fig:tradeoff}
\end{figure}

%
%

  Based on the enhanced VIP algorithms in Algorithm~\ref{Alg:VIP} and Algorithm~\ref{Alg:flow-DBP}  operating on VIPs in the virtual plane, we can develop the corresponding algorithms for handling  Interest Packets and Data Packets in the actual plane using a mapping similar to that in \cite{VIPICN14,VIPfull16}. We omit the details due to the page limitation. In this section, we first compare the delay  performance of  the   enhanced VIP algorithm for the actual plane  resulting from Algorithm~\ref{Alg:VIP}   (using  the minimum next-hop bias function in \eqref{eqn:ex-onehop-f} with $z=1$), denoted by EVIP,  with the VIP algorithm for the actual plane resulting from Algorithm~1 in \cite{VIPICN14,VIPfull16}, denoted by VIP, as well as with six other baseline algorithms. In particular, these baseline algorithms use popular caching algorithms (LFU, LCE-UNIF, LCE-LRU, LCD-LRU, and LCE-BIAS)  in conjunction with shortest path forwarding and a potential-based forwarding algorithm. The detailed descriptions of these baseline algorithms can be found  in \cite{VIPfull16}.  We then compare the utility-delay tradeoff of  the enhanced VIP algorithm for the actual plane resulting from Algorithm~\ref{Alg:flow-DBP} (using  the minimum next-hop bias function in \eqref{eqn:ex-onehop-f} with $z=1$), also denoted by EVIP,  with the VIP algorithm for the actual plane resulting from Algorithm~3 in \cite{VIPfull16}, also  denoted by VIP, with some abuse of notation. To our knowledge, there are no other congestion control algorithms for NDN networks which can easily control the tradeoff between utility and delay. In  evaluating delay and utility-delay tradeoff, as in \cite{VIPICN14,VIPfull16}, we also consider the constant shortest path bias  versions of EVIP and VIP, with $B$ being the per-link cost.

 Experimental scenarios are carried on two network topologies:  the GEANT Topology  and the  DTelekom Topology, as shown in Fig.~\ref{Fig:topo}. In the two topologies, object requests can be generated by any node, and the content source for each data object is independently and uniformly distributed among all nodes. At each node requesting data, object requests arrive according to a Poisson process  with an overall rate $\lambda$ (in requests/node/slot). Each arriving request requests data object $k$ (independently) with probability $p_k$, where $\{p_k\}$ follows a (normalized) Zipf distribution with parameter 0.75.
We choose $K=3000$, $C_{ab}=500$ Mb/slot, $L_n=2$ GB, $D=5$ MB, the Interest Packet size is 125B, and the Data Packet size is 50 KB.  Each simulation generates requests  for  $10^4$ time slots.  Each curve   is obtained by averaging over 10 simulation runs.
The delay  for an Interest Packet request is the difference  (in time slots) between the fulfillment
time (i.e., time of arrival of the requested Data Packet) and the creation time
of the Interest Packet request. We use the total delay for all the Interest Packets generated over $10^4$ time slots as the delay   measure. We consider $\alpha$-utility function with $\alpha=2$, i.e., $g_n^k(x)=-\frac{1}{x}$, and use the sum utility over all nodes and all objects as the utility  measure.

 Fig.~\ref{Fig:delay} illustrates the delay performance. We can observe that the VIP algorithms  in \cite{VIPICN14,VIPfull16}  and the proposed enhanced VIP algorithms  achieve much better delay performance than the six baseline schemes, especially in DTelekom. In addition, the proposed enhanced VIP algorithms   significantly improve the  delay performance of the VIP algorithms (e.g., about $28\%$ at   $\lambda=30$ in GEANT and  $30\%$ at  $\lambda=40$ in DTelekom).  Fig.~\ref{Fig:tradeoff} illiterates the utility-delay tradeoff. We can observe that the proposed enhanced VIP algorithms  achieve significantly better utility-delay tradeoff than the VIP algorithms in \cite{VIPfull16}. In summary, the proposed enhanced VIP algorithms improve the delay performance of  the VIP algorithms in \cite{VIPICN14,VIPfull16} by intelligently exploiting the VIP counts beyond one hop for forwarding, caching and congestion control.

\section{Conclusion}

 In this paper,
we develop a new class of enhanced distributed  forwarding, caching and congestion control algorithms within the VIP framework, which adaptively stabilize the network and maximize network utility, while improving delay performance.  We prove the throughput optimality and characterize the utility-delay tradeoff of the enhanced VIP algorithms  in the virtual  plane.  Numerical experiments  demonstrate the superior performance of the resulting algorithms  for handling  Interest Packets and Data Packets  in the actual plane, in terms of low network delay and high network utility,  relative to  a number of baseline alternatives.

\newpage

\section*{Appendix A: Proof of Theorem~\ref{Thm:thpt-opt}}

Define the quadratic Lyapunov function $\mathcal L(\mathbf V)\triangleq
\sum_{n\in \mathcal N,k\in \mathcal K}(V^k_n)^2$. The Lyapunov drift at slot $t$ is given by
$\Delta (\mathbf V(t))\triangleq \mathbb E[\mathcal L\big(\mathbf
V(t+1)\big)-\mathcal L\big(\mathbf V(t)\big)|\mathbf V(t)]$. First, we calculate $\Delta (\mathbf V(t))$.
Taking square on both sides of \eqref{eqn:queue_dyn}, we have
\begin{align}
&\left(V^k_n(t+1)\right)^2\nonumber\\
\leq&\Bigg(\Bigg(
\left(V^k_n(t)-\sum_{b\in \mathcal N}\mu^k_{nb}(t)\right)^+ +A^k_n(t)
\nonumber\\
&+\sum_{a\in \mathcal N}\mu^k_{an}(t)-r_ns_n^k(t)\Bigg)^+\Bigg)^2\nonumber\\
\leq&\Bigg(
\left(V^k_n(t)-\sum_{b\in \mathcal N}\mu^k_{nb}(t)\right)^+ +A^k_n(t)\nonumber\\
&+\sum_{a\in \mathcal N}\mu^k_{an}(t)-r_ns_n^k(t)\Bigg)^2\nonumber\\
\leq&\left(V^k_n(t)-\sum_{b\in \mathcal N}\mu^k_{nb}(t)\right)^2+2
\left(V^k_n(t)-\sum_{b\in \mathcal N}\mu^k_{nb}(t)\right)^+\nonumber\\
&\times\left(A^k_n(t)
+\sum_{a\in \mathcal N}\mu^k_{an}(t)-r_ns_n^k(t)\right)\nonumber\\
&+\left(A^k_n(t)
+\sum_{a\in \mathcal N}\mu^k_{an}(t)-r_ns_n^k(t)\right)^2
\nonumber\\
=&\left(V^k_n(t)\right)^2+\left(\sum_{b\in \mathcal N}\mu^k_{nb}(t)\right)^2-2 V^k_n(t)\sum_{b\in \mathcal N}\mu^k_{nb}(t) \nonumber\\
&+\left(A^k_n(t)+\sum_{a\in \mathcal N}\mu^k_{an}(t)-r_n^ks_n^k(t)\right)^2\nonumber\\
&+2\left(V^k_n(t)-\sum_{b\in \mathcal N}\mu^k_{nb}(t)\right)^+\left(A^k_n(t)
+\sum_{a\in \mathcal N}\mu^k_{an}(t)\right)\nonumber\\
&-2\left(V^k_n(t)-\sum_{b\in \mathcal N}\mu^k_{nb}(t)\right)^+r_n^ks_n^k(t)\nonumber\\
\leq&\left(V^k_n(t)\right)^2+\left(\sum_{b\in \mathcal N}\mu^k_{nb}(t)\right)^2-2 V^k_n(t)\sum_{b\in \mathcal N}\mu^k_{nb}(t) \nonumber\\
&+\left(A^k_n(t)+\sum_{a\in \mathcal N}\mu^k_{an}(t)+r_n^ks_n^k(t)\right)^2\nonumber\\
&+2V^k_n(t)\left(A^k_n(t)
+\sum_{a\in \mathcal N}\mu^k_{an}(t)\right)\nonumber\\
&-2\left(V^k_n(t)-\sum_{b\in \mathcal N}\mu^k_{nb}(t)\right)r_n^ks_n^k(t)\nonumber
\end{align}
\begin{align}
\leq&\left(V^k_n(t)\right)^2+\left(\sum_{b\in \mathcal N}\mu^k_{nb}(t)\right)^2+2\sum_{b\in \mathcal N}\mu^k_{nb}(t)r_ns_n^k(t)\nonumber\\
&+\left(A^k_n(t)+\sum_{a\in \mathcal N}\mu^k_{an}(t)+r_ns_n^k(t)\right)^2\nonumber\\
&+2V^k_n(t)A^k_n(t)-2V^k_n(t)\left(\sum_{b\in \mathcal N}\mu^k_{nb}(t)-\sum_{a\in \mathcal N}\mu^k_{an}(t)\right)\nonumber\\
&-2V^k_n(t)r_ns_n^k(t)\nonumber
\end{align}
Summing over all $n,k$, we have
\begin{align}
&\mathcal L\left(\mathbf V(t+1)\right)-\mathcal L\left(\mathbf V(t)\right)\nonumber\\
\stackrel{(a)}{\leq}&2N B+2\sum_{n\in \mathcal N,k\in \mathcal K}V^k_n(t)A^k_n(t)\nonumber\\
&-2\sum_{(a,b)\in \mathcal L}\sum_{k\in \mathcal K}
\mu^k_{ab}(t)\big(V^k_a(t)+f_a^k(\mathbf V(t))-V^k_b(t)-f_b^k(\mathbf V(t))\big)\nonumber\\
&+2\sum_{(a,b)\in \mathcal L}\sum_{k\in \mathcal K}
\mu^k_{ab}(t)\big(f^k_a(\mathbf V(t))-f^k_b(\mathbf V(t))\big)\nonumber\\
&-2\sum_{n\in \mathcal N,k\in \mathcal K}\big(V^k_n(t)+f_n^k(\mathbf V(t))\big)r_ns_n^k(t)\nonumber\\
&+2\sum_{n\in \mathcal{N},k\in \mathcal K}f_n^k(\mathbf V(t))r_ns_n^k(t) \nonumber\\
\stackrel{(b)}{\leq}&2N B+2\sum_{n\in \mathcal N,k\in \mathcal K}V^k_n(t)A^k_n(t)\nonumber\\
&-2\sum_{(a,b)\in \mathcal L}\sum_{k\in \mathcal K}
\mu^k_{ab}(t)\big(V^k_a(t)+f_a^k(\mathbf V(t))-V^k_b(t)-f_b^k(\mathbf V(t))\big)\nonumber\\
&-2\sum_{n\in \mathcal N,k\in \mathcal K}\big(V_n^k(t)+f_n^k(\mathbf V(t))\big)r_ns_n^k(t) \nonumber\\
&+2\sum_{n\in \mathcal N,k\in \mathcal K}\frac{V_n^k(t)}{z_n^k}\left(C_{\max}L^k+Nr_{\max}\right)
\label{eqn:proof-deltaL}
\end{align}
where (a) is due to the following:
\begin{align}
&\sum_{k\in \mathcal K}\left(\sum_{b\in \mathcal N}\mu^k_{nb}(t)\right)^2\leq \left(\sum_{k\in \mathcal K}\sum_{b\in \mathcal N}\mu^k_{nb}(t)\right)^2\leq\left(\mu^{out}_{n, \max}\right)^2,\nonumber\\
& \sum_{k\in \mathcal K}\left(A^k_n(t)+\sum_{a\in \mathcal N}\mu^k_{an}(t)+r_ns_n^k(t)\right)^2\nonumber\\
&\leq \left(\sum_{k\in \mathcal K}\left(A^k_n(t)+\sum_{a\in \mathcal N}\mu^k_{an}(t)+r_ns_n^k(t)\right)\right)^2\nonumber\\
 &\leq (A_{n,\max}+\mu^{in}_{n,\max}+Kr_n)^2, \nonumber
 \end{align}
 \begin{align}
&\sum_{k\in \mathcal K}\sum_{b\in \mathcal N}\mu^k_{nb}(t)r_ns_n^k(t)\nonumber\\
&\leq \left(\sum_{k\in \mathcal K}\sum_{b\in \mathcal N}\mu^k_{nb}(t)\right)\left(\sum_{k\in \mathcal K}r_ns_n^k(t)\right)\leq \mu^{out}_{n,
\max}Kr_{n},\nonumber
\end{align}
\begin{align}
&\sum_{n\in \mathcal N,k\in \mathcal K}V^k_n(t)\left(\sum_{b\in\mathcal N}\mu^k_{nb}(t)-\sum_{a\in \mathcal N}\mu^k_{an}(t)\right)\nonumber\\
&=\sum_{(a,b)\in \mathcal L}\sum_{k\in \mathcal K}
\mu^k_{ab}(t)\big(V^k_a(t)-V^k_b(t)\big).\nonumber
\end{align}
and (b) is due to the following:
\begin{align}
&\sum_{(a,b)\in \mathcal L}\sum_{k\in \mathcal K}
\mu^k_{ab}(t)\big(f_a^k(\mathbf V(t))-f_b^k(\mathbf V(t))\big) \nonumber\\
&=\sum_{(a,b)\in \mathcal L}\sum_{k\in \mathcal K}
\mu^k_{ab}(t)\sum_{n\in \mathcal N}\frac{\eta_{an}^k(\mathbf V(t))-\eta_{bn}^k(\mathbf V(t))}{z_n^k}V_n^k(t) \nonumber\\
&\leq \sum_{k\in \mathcal K}\sum_{(a,b)\in \mathcal L^k} C_{\max}\sum_{n\in \mathcal N}\frac{1}{z_n^k}V_n^k(t) \nonumber\\
&=\sum_{n\in \mathcal N,k\in \mathcal K}V_n^k(t)\frac{C_{\max}L^k}{z_n^k}\label{eqn:proof-fa-fb}\\
&\sum_{n\in \mathcal{N},k\in \mathcal K}f_n^k(\mathbf V(t))r_ns_n^k(t)\nonumber\\
\leq &\sum_{n\in \mathcal{N},k\in \mathcal K}\left(\sum_{n'\in\mathcal N}\frac{1}{z_{n'}^k}V_{n'}^k(t)\right)r_{\max}\nonumber\\
=&\sum_{n\in \mathcal{N},k\in \mathcal K}V_{n}^k(t)\frac{Nr_{\max}}{z_{n}^k}\nonumber
\end{align}
Taking conditional
expectations on both sides of \eqref{eqn:proof-deltaL}, we have
\begin{align}
&\Delta (\mathbf
V(t))\nonumber\\
\leq&2N B+2\sum_{n\in \mathcal N,k\in \mathcal K}V^k_n(t)\lambda^k_n\nonumber\\
&-2\mathbb
E\bigg[\sum_{(a,b)\in \mathcal L}\sum_{k\in \mathcal K}
\mu^k_{ab}(t)\bigg(((V^k_a(t)+f_a^k(t))\nonumber \\
&\hspace{40mm} -(V^{k}_b(t)+f_b^k(t))\bigg)|\mathbf V(t)\bigg]\nonumber\\
&-2\mathbb E\left[\sum_{n\in \mathcal N,k\in \mathcal K}\big(V^k_n(t)+f_n^k(t)\big)r_n s_n^k(t)|\mathbf V(t)\right]\nonumber\\
&+2\sum_{n\in \mathcal N,k\in \mathcal K}\frac{V_n^k(t)}{z_n^k}\left(C_{\max}L^k+Nr_{\max}\right)\nonumber\\
\stackrel{(c)}{\leq}&2N B+2\sum_{n\in \mathcal N,k\in \mathcal K}V^k_n(t)\lambda^k_n\nonumber\\
&-2\mathbb
E\bigg[\sum_{(a,b)\in \mathcal L}\sum_{k\in \mathcal K}
\tilde{\mu}^k_{ab}(t)\bigg((V^k_a(t)+f_a^k(t)) \nonumber\\
&-(V^k_b(t)+f_b^k(t))\bigg)|\mathbf V(t)\bigg]\nonumber\\
&-2\mathbb E\left[\sum_{n\in \mathcal N,k\in \mathcal K}\big(V^k_n(t)+f_n^k(t)\big)r_n\tilde{s}_n^k(t)|\mathbf V(t)\right]\nonumber
\end{align}
\begin{align}
&+2\sum_{n\in \mathcal N,k\in \mathcal K}\frac{V_n^k(t)}{z_n^k}\left(C_{\max}L^k+Nr_{\max}\right)\nonumber\\
\leq&2N B+2\sum_{n\in \mathcal N,k\in \mathcal K}V^k_n(t)\lambda^k_n-2\sum_{n\in \mathcal N,k\in \mathcal K}V^k_n(t)\nonumber\\
&\times\mathbb E\left[\left(\sum_{b\in \mathcal N}\tilde{\mu}^k_{nb}(t)-\sum_{a\in \mathcal N}\tilde{\mu}^k_{an}(t)+r_n\tilde{s}_n^k(t)\right)|\mathbf V(t)\right] \nonumber \\
&+2\mathbb
E\bigg[\sum_{(a,b)\in \mathcal L}\sum_{k\in \mathcal K}
\tilde{\mu}^k_{ab}(t)\left(f_b^k(t)-f_a^k(t)\right)|\mathbf V(t)\bigg]\nonumber\\
&-2\mathbb E\left[\sum_{n\in \mathcal N,k\in \mathcal K} f_n^k(t)r_n\tilde{s}_n^k(t)|\mathbf V(t)\right]\nonumber\\
&+2\sum_{n\in \mathcal  N,k \in \mathcal K}\frac{V_n^k(t)}{z_n^k}\big(C_{\max}L^k+Nr_{max}\big)\label{eqn:proof_ineq0}
\end{align}
where (c) is due to the fact that Algorithm \ref{Alg:VIP} minimizes the
R.H.S. of  (c) over all feasible $\tilde{\mu}^k_{ab}(t)$ and $\tilde{s}_n^k(t)$.\footnote{ Note that $\mu^k_{ab}(t)$ and $s_n^k(t)$ denote the actions of Algorithm \ref{Alg:VIP}.}  Since $\boldsymbol
\lambda+\boldsymbol \epsilon + \boldsymbol \delta \in \Lambda$, according to the proof of Theorem~1 in \cite{VIPfull16}, there exists a stationary randomized forwarding and caching policy that makes decisions
independent of $\mathbf V(t)$ such that
\begin{align}
&\mathbb E\left[\left(\sum_{b\in \mathcal N}\tilde{\mu}^k_{nb}(t)-\sum_{a\in \mathcal N}\tilde{\mu}^k_{an}(t)+r_n\tilde{s}_n^k(t)\right)|\mathbf
V(t)\right]\nonumber\\
= &\lambda^k_n+\epsilon^k_n+\delta_n^k\label{eqn:proof_ineq1}
\end{align}
 On the other hand, similar to \eqref{eqn:proof-fa-fb}, we can show
\begin{align}
&\mathbb
E\bigg[\sum_{(a,b)\in \mathcal L}\sum_{k\in \mathcal K}
\tilde{\mu}^k_{ab}(t)\left(f_b^k(t)-f_a^k(t)\right)|\mathbf V(t)\bigg]\nonumber\\
\leq&\sum_{n\in \mathcal N,k\in \mathcal K}V_n^k(t)\frac{C_{\max}L^k}{z_n^k}\label{eqn:proof-fb-fa}
\end{align}
By \eqref{eqn:proof_ineq0}, \eqref{eqn:proof-fb-fa} and $\tilde{s}_n^k(t)\geq 0$ for all $n$ and $k$, we have  $\Delta (\mathbf V(t))
\leq 2NB-2\min_{n\in \mathcal N,k\in \mathcal K}\left\{\epsilon^k_n+\delta_n^k -\frac{2C_{\max}L^k+Nr_{\max}}{z_n^k}\right\}
\sum_{n \in N,k\in K}V_n^k(t)$.
By  Lemma 4.1 of \cite{Georgiadis-Neely-Tassiulas:2006}, we complete the proof.

\section*{Appendix B: Proof of Theorem~\ref{Thm:flow-control}}

Define the Lyapunov function $\mathcal L(\boldsymbol \Theta)\triangleq
\sum_{n\in \mathcal N,k\in \mathcal K}\left((V^k_n)^2+(Y^k_n)^2\right)$, where $\boldsymbol \Theta\triangleq (\mathbf V, \mathbf Y)$. The Lyapunov drift at slot $t$ is
$\Delta (\boldsymbol \Theta(t))\triangleq \mathbb E[\mathcal L\big(\boldsymbol \Theta(t+1)\big)- \mathcal  L\left(\boldsymbol \Theta(t)\right)|\boldsymbol \Theta(t)]$.  First, we calculate $\Delta (\boldsymbol \Theta(t))$. Similar to Appendix A, taking square on both sides of \eqref{eqn:queue_dyn-flow}, we have
\begin{align}
&\left(V^k_n(t+1)\right)^2\nonumber\\
\leq&\left(V^k_n(t)\right)^2+\left(\sum_{b\in \mathcal N}\mu^k_{nb}(t)\right)^2+2\sum_{b\in \mathcal N}\mu^k_{nb}(t)r_ns_n^k(t)\nonumber\\
&+\left(\alpha^k_n(t)+\sum_{a\in \mathcal N}\mu^k_{an}(t)+r_ns_n^k(t)\right)^2\nonumber\\
&+2V^k_n(t)\alpha^k_n(t)-2V^k_n(t)\left(\sum_{b\in \mathcal N}\mu^k_{nb}(t)-\sum_{a\in \mathcal N}\mu^k_{an}(t)\right)\nonumber\\
&-2V^k_n(t)r_ns_n^k(t)\nonumber
\end{align}
In addition, taking square on both sides of \eqref{eqn:queue_dyn-virtual}, we have
\begin{align}
&\left(Y^k_n(t+1)\right)^2\nonumber\\
\leq&\left(Y^k_n(t)\right)^2+\left(\alpha^k_n(t)\right)^2+\left(\gamma^k_n(t)\right)^2 -2Y_n^k(t)\left(\alpha_n^k(t)-\gamma_n^k(t)\right)\nonumber
\end{align}
Therefore, similarly, we have
\begin{align}
&\mathcal  L\left(\boldsymbol \Theta (t+1)\right)- \mathcal  L\left(\boldsymbol \Theta (t)\right)\nonumber\\
 \leq& 2N\hat B+2\sum_{n\in \mathcal N,k\in \mathcal K}V^k_n(t)\alpha^k_n(t)\nonumber\\
&  -2\sum_{(a,b)\in \mathcal L,k\in \mathcal K}\mu_{ab}^k(t)\Big((V_a^k(t)+f_a^k(\mathbf V(t)))\nonumber\\
&\hspace{30mm}-(V_b^k(t)+f_b^k(\mathbf V(t)))\Big)\nonumber\\
& +2\sum_{(a,b)\in \mathcal L,k\in \mathcal K}\mu_{ab}^k(t)\left(f_a^k(\mathbf V(t))-f_b^k(\mathbf V(t))\right)\nonumber\\
& -2\sum_{n\in \mathcal N,k\in \mathcal K}\big(V^k_n(t)+f_n^k(\mathbf V(t))\big)r_ns_n^k(t)\nonumber\\
& +2\sum_{n\in \mathcal N,k\in \mathcal K}f_n^k(\mathbf V(t))r_ns_n^k(t)\nonumber\\
& -2\sum_{n\in \mathcal N,k\in \mathcal K}Y_n^k(t)\left(\alpha_n^k(t)-\gamma_n^k(t)\right) \nonumber\\
 \leq& 2N\hat B+2\sum_{n\in \mathcal N,k\in \mathcal K}V^k_n(t)\alpha^k_n(t)\nonumber\\
&  -2\sum_{(a,b)\in \mathcal L,k\in \mathcal K}\mu_{ab}^k(t)\Big((V_a^k(t)+f_a^k(\mathbf V(t)))\nonumber\\
&\hspace{30mm}-(V_b^k(t)+f_b^k(\mathbf V(t)))\Big)\nonumber\\
& -2\sum_{n\in \mathcal N,k\in \mathcal K}\big(V^k_n(t)+f_n^k(\mathbf V(t))\big)r_ns_n^k(t)\nonumber\\
& -2\sum_{n\in \mathcal N,k\in \mathcal K}Y_n^k(t)\left(\alpha_n^k(t)-\gamma_n^k(t)\right) \nonumber\\
& +2\sum_{n\in \mathcal N,k\in \mathcal K}\frac{V_n^k(t)}{z_n^k}\big(C_{\max}L^k+Nr_{\max}\big)
\label{eqn:proof-drift-flow}
\end{align}
Taking conditional
expectations and subtracting $$W\mathbb E\left[\sum_{n\in \mathcal N,k\in \mathcal K}g^k_n\left( \gamma^k_n(t)\right)|\boldsymbol \Theta (t)\right]$$ from both sides of \eqref{eqn:proof-drift-flow}, we have
\begin{align}
&\Delta \left(\boldsymbol \Theta (t)\right )-W\mathbb E\left[\sum_{n\in \mathcal N,k\in \mathcal K}g^k_n\left( \gamma^k_n(t)\right)|\boldsymbol \Theta(t)\right]\nonumber\\
 \stackrel{(a)}{\leq}& 2N\hat B-2\sum_{n\in \mathcal N,k\in \mathcal K}\left(Y^k_n(t)-V^k_n(t)\right)\mathbb E\left[\tilde \alpha^k_n(t)|\boldsymbol \Theta(t)\right]\nonumber\\
& -\sum_{n\in \mathcal N,k\in \mathcal K}\mathbb E\left[Wg^k_n\left( \tilde \gamma^k_n(t)\right)-2Y_n^k(t)\tilde \gamma^k_n(t)|\boldsymbol \Theta(t)\right]\nonumber\\
&  - 2\sum_{n\in \mathcal N,k\in \mathcal K}V^k_n(t)\nonumber\\
&\times\mathbb E\left[\left(\sum_{b\in \mathcal N}\tilde{\mu}^k_{nb}(t)-\sum_{a\in \mathcal N}\tilde{\mu}^k_{an}(t)+r_n\tilde{s}_n^k(t)\right)|\boldsymbol \Theta(t)\right] \nonumber \\
&+2\sum_{(a,b)\in \mathcal L}\sum_{k\in \mathcal K}\mathbb E\left[\left(\tilde{\mu}_{ab}(t)
\big(f_b^k(\mathbf V(t))- f_a^k(\mathbf V(t))\big)|\boldsymbol \Theta(t)\right)\right] \nonumber\\
&-2\sum_{n\in \mathcal N,k\in \mathcal K}\mathbb E\left[f_n^k(\mathbf V(t))r_n\tilde{s_n}_n^k(t)|\boldsymbol \Theta(t)\right] \nonumber\\
\leq& 2N\hat B-2\sum_{n\in \mathcal N,k\in \mathcal K}\left(Y^k_n(t)-V^k_n(t)\right)\mathbb E\left[\tilde \alpha^k_n(t)|\boldsymbol \Theta(t)\right]\nonumber\\
& -\sum_{n\in \mathcal N,k\in \mathcal K}\mathbb E\left[Wg^k_n\left( \tilde \gamma^k_n(t)\right)-2Y_n^k(t)\tilde \gamma^k_n(t)|\boldsymbol \Theta(t)\right]\nonumber\\
&  - 2\sum_{n\in \mathcal N,k\in \mathcal K}V^k_n(t)\nonumber\\
&\times\mathbb E\left[\left(\sum_{b\in \mathcal N}\tilde{\mu}^k_{nb}(t)-\sum_{a\in \mathcal N}\tilde{\mu}^k_{an}(t)+r_n\tilde{s}_n^k(t)\right)|\boldsymbol \Theta(t)\right] \nonumber \\
&+2\sum_{n\in \mathcal N,k\in \mathcal K}\frac{V_n^k(t)}{z_n^k}(2R_{max}L^k+Nr_{max})
\label{eqn:proof_ineq0-flow}
\end{align}
where (a) is due to the fact that  Algorithm \ref{Alg:flow-DBP} minimizes the
R.H.S. of  (b) over all possible alternative
$\tilde \alpha^k_n (t)$, $\tilde \gamma^k_n(t)$, $\tilde \mu^k_{ab}(t)$, and $\tilde s^k_n(t)$.\footnote{Note that $\alpha^k_n (t)$, $\gamma^k_n(t)$, $\mu^k_{ab}(t)$ and $s^k_n(t)$
denote the actions of Algorithm \ref{Alg:flow-DBP}.}
It is not difficult to construct alternative random policies that choose $\tilde \alpha^k_n (t)$, $\tilde \gamma^k_n(t)$, $\tilde \mu^k_{ab}(t)$ and $\tilde s^k_n(t)$ such that
\begin{align}
&\mathbb E\left[\tilde \alpha^k_n(t)|\boldsymbol \Theta(t)\right]=\overline \alpha^{k*}_n(\boldsymbol \epsilon+\boldsymbol \delta)\label{eqn:rand-r}\\
& \tilde \gamma^k_n(t)=\overline \alpha^{k*}_n(\boldsymbol \epsilon+\boldsymbol \delta)\label{eqn:rand-gamma}\\
& \mathbb E\left[\left(\sum_{b\in \mathcal N}\tilde{\mu}^k_{nb}(t)-\sum_{a\in \mathcal N}\tilde{\mu}^k_{an}(t)+r_n\tilde{s}_n^k(t)\right)|\boldsymbol \Theta(t)\right]\nonumber\\
=& \overline \alpha^{k*}_n(\boldsymbol \epsilon+\boldsymbol \delta)+\epsilon^k_n + \delta_n^k\label{eqn:rand-mu}
\end{align}
where  $ \overline{\boldsymbol \alpha}^*(\boldsymbol \epsilon+\boldsymbol \delta)=(\overline \alpha^{k*}_n(\boldsymbol \epsilon+\boldsymbol \delta))$ is the target $\boldsymbol \epsilon $-optimal admitted rate given by
\eqref{eqn:eps-opt-prob}.\footnote{Specifically, \eqref{eqn:rand-r} can be achieved by the random policy setting $\tilde \alpha^k_n(t)=A^k_n(t)$ with probability $\overline \alpha^{k*}_n(\boldsymbol \epsilon)/\lambda^k_n$ and $\tilde \alpha^k_n(t)=0$ with probability $1-\overline \alpha^{k*}_n(\boldsymbol \epsilon)/\lambda^k_n$.}
\eqref{eqn:rand-mu} follows from the same arguments leading to~\eqref{eqn:proof_ineq1}.
Thus, by \eqref{eqn:rand-r}, \eqref{eqn:rand-gamma} and \eqref{eqn:rand-mu}, from \eqref{eqn:proof_ineq0-flow}, we obtain
\begin{align}
&\Delta (\boldsymbol \Theta (t))-W\mathbb E\left[\sum_{n\in \mathcal N,k\in \mathcal K}g^k_n\left( \gamma^k_n(t)\right)|\boldsymbol \Theta(t)\right]\nonumber\\
 \leq &  2N\hat B-2\min_{n\in \mathcal N,k\in \mathcal K}\left\{\epsilon^k_n+\delta_n^k -\frac{2C_{\max}L^k+Nr_{\max}}{z_n^k}\right\} \nonumber\\
 & \times\sum_{n\in \mathcal N,k\in \mathcal K}V_n^k(t)-W\sum_{n\in \mathcal N,k\in \mathcal K}g^k_n\left(\overline \alpha^{k*}_n(\boldsymbol \epsilon +\boldsymbol \delta)\right)\nonumber
\end{align}
Applying Theorem 5.4 of \cite{Georgiadis-Neely-Tassiulas:2006}, we have
\begin{align}
&\limsup_{t\to\infty}\frac{1}{t}\sum_{\tau=1}^t\sum_{n\in \mathcal N,k\in \mathcal K} \mathbb
E[V^k_n(\tau)]  \nonumber\\
\leq& \frac{2N\hat B+WG_{\max}}{
2\min_{n\in \mathcal N,k\in \mathcal K}\left\{\epsilon^k_n+\delta_n^k -\frac{2C_{\max}L^k+Nr_{\max}}{z_n^k}\right\}}\label{eqn:proof-enhanced-flow-DBP-U}\\
&\liminf_{t\to\infty}\sum_{n\in \mathcal N,k\in \mathcal K} g^k_n\left(\overline \gamma^k_n(t)\right)\nonumber\\
  \geq&
\sum_{n\in \mathcal N,k\in \mathcal K}
g^k_n\left( \overline \alpha^{k*}_n\left(\boldsymbol \epsilon+\boldsymbol \delta\right)\right)-\frac{2N\hat B}{W}\label{eqn:proof-enhanced-flow-DBP-g}
\end{align}
As in \cite[page 88]{Georgiadis-Neely-Tassiulas:2006}, we optimize the R.H.S. of  \eqref{eqn:proof-enhanced-flow-DBP-U} and \eqref{eqn:proof-enhanced-flow-DBP-g} over all possible $(\boldsymbol \epsilon,\boldsymbol\delta) $. Thus, we can show \eqref{eqn:enhanced-flow-DBP-U} and
\begin{align}
& \liminf_{t\to\infty}\sum_{n\in \mathcal N,k\in \mathcal K}g^k_n\left(\overline \gamma^k_n(t)\right) \nonumber\\
\geq&
\sum_{n\in \mathcal N,k\in \mathcal K}
g^k_n\left(\overline \alpha^{k*}_n\left(\mathbf 0\right)\right)-\frac{2N\hat B}{W}
\label{eqn:enhanced-flow-DBP-g-sim}
\end{align}
where $\overline \gamma^k_n(t)\triangleq \frac{1}{t}\sum_{\tau=1}^t\mathbb E[ \gamma^k_n(\tau)]$.
It is easy to prove $\overline \gamma^k_n(t)\leq \overline \alpha^k_n(t)$ by showing the stability of the virtual queues. Thus, we can show \eqref{eqn:enhanced-flow-DBP-g} based on \eqref{eqn:enhanced-flow-DBP-g-sim}. We complete the proof.

\end{document}